# WIND-Bench: A Benchmark Dataset for In-Situ Near-Surface Wind Speed Observations Across the Conterminous United States

**Authors**
Kyla Bazlen[1], Grant Buster[2*], Brandon Benton[2], Lauren North[1], Ansley Baring[3], David D. Turner[3], Emily Wells[4], Laura Vimmerstedt[2].

**Affiliations**
1. NSF ASCEND Engine, Fort Collins, Colorado, USA
2. Strategic Energy Analysis Center, National Laboratory of the Rockies, Golden, Colorado, USA
3. Global Systems Laboratory, National Oceanic and Atmospheric Administration, Boulder, Colorado, USA
4. Cooperative Institute for Research in the Atmosphere, Fort Collins, Colorado, USA

*Corresponding author: Grant Buster (Grant.Buster@nlr.gov)

# Abstract

Accurate wind forecasts are essential for operational decision-making and public safety, yet forecasts tend to miss near-surface high wind speeds in complex terrain. In response, advances in machine learning (ML) weather prediction methods have demonstrated the ability to improve forecast skill beyond traditional numerical weather prediction (NWP) models. However, the absence of a benchmark dataset to evaluate NWP and ML models with sufficient, quality-controlled wind speed observations in complex terrain poses challenges to the development and intercomparison of high-quality surface wind forecasts across the Conterminous United States (CONUS). We develop the Wind IN-situ Data Benchmark (WIND-Bench), a benchmark dataset from in-situ observations in the Meteorological Assimilation Data Ingest System (MADIS) observational network. WIND-Bench integrates multiple sensor networks with quality control that distinguishes sensor failures from high-wind conditions, using a framework that validates observations against forecasts from the National Oceanic and Atmospheric Administration (NOAA) High-Resolution Rapid Refresh (HRRR) model. WIND-Bench provides a standardized benchmark for evaluating ML and NWP models and for quantifying forecast skill, accelerating the development, evaluation, and operational deployment of skilled near-surface wind forecasts.

# Introduction and Summary

High-quality weather forecasts are essential for public safety, emergency and operational decision-making, and economic savings (Bauer et al. 2015; Turner et al. 2022). Weather forecasts primarily rely on numerical weather prediction (NWP) models, which have advanced substantially over the past century (Bauer et al. 2015; Benjamin et al. 2018). Accurate wind forecasts are particularly critical for mitigating impacts from hazardous weather events, including wildfires, hurricanes, and tornadoes, by enabling timely public warnings and improving response preparedness (Molina and Rudik 2024; Taylor et al. 2018). In wildfire risk management, electric utilities rely on wind forecasts in combination with relative humidity and temperature to inform Public Safety Power Shutoffs (PSPS), where distribution and/or transmission lines are de-energized to prevent ignitions during times of dangerous fire weather (Fovell and Capps 2024). Forecast accuracy in these instances has societal consequences, as both implemented and canceled PSPS events can cause harm to public health and lead to economic losses (Wong-Parodi 2020; Huang et al. 2023). With hot, dry extreme events increasing across CONUS, and extreme fire weather days projected to increase, well-validated short- and medium-range wind forecasts are increasingly essential to public safety and economic well-being across the country (Gamelin et al. 2022; Alizadeh et al. 2020; Abatzoglou et al. 2019).

High-wind events in complex terrain are particularly challenging to forecast because of the wind's spatial heterogeneity and its complex interactions with topography, which create phenomena such as gap flow, mountain waves, and downslope flow (Collins et al. 2024a). The High-Resolution Rapid Refresh Model (HRRR), the highest-resolution operational model in the United States at a 3km resolution, overestimates near-surface wind speeds in forested regions and underestimates extreme winds in mountainous terrain (Y. Liu et al. 2025). HRRR underpredicts approximately 99% of winds exceeding 20 m $s^{-1}$ across the Colorado-Wyoming region, exemplifying current limitations in forecasting severe wind events (Collins et al. 2024a).

These forecasting challenges for high wind and complex terrain are consequential because wildfire risk is already elevated under these conditions; slopes of greater than 30° increase heat release and spread rates and limit the accessibility of responding fire crews (Kim et al. 2025; Werth et al. 2011; Linn et al. 2007; Hayajneh and Naser 2025). Therefore, improving wind forecasts in these conditions remains an important challenge.

In recent years, there has been a rapid rise in the use of machine learning (ML) and artificial intelligence (AI) in Earth systems modeling. In some cases, ML approaches outperform traditional physics-based NWP models and have shown promise for improving wind forecasting in complex terrain by capturing non-linear patterns in local atmospheric dynamics that are difficult to parameterize in NWPs (Goutham et al. 2021; D. Liu et al. 2025; Valdivia-Bautista et al. 2023). Forecasting using ML models is also gaining traction for its computational efficiency relative to NWP models, making forecasts 10,000 to 100,000 times faster (Bi et al. 2023; Pathak et al. 2022). To train and evaluate the skill of ML forecasts, high-quality benchmark datasets are needed.

Benchmark datasets enable standardized, reproducible evaluation and validation for weather models (Dueben et al. 2022). In response to the proliferation of AI weather models, several have been developed, including the global benchmarks WeatherBench2 and WeatherReal (Jin et al. 2024; Rasp et al. 2024). WeatherBench2 uses ERA5 and the Integrated Forecasting System (IFS) High Resolution (HRES) forecasts as “ground-truth datasets”, supplemented by observation data from the Meteorological Assimilation Data Ingest System (MADIS) METeorological Aerodrome Report (METAR) network (Rasp et al. 2024). This dataset has been used widely by the meteorology community, enabling the development of well-performing global ML models such as GraphCast and Aurora (Bodnar et al. 2025; Lam et al. 2023). While ERA5 provides consistent hourly global coverage and has been extensively validated, it exhibits systematic errors in wind speed in coastal regions and complex terrain (Gualtieri 2021; Wilczak et al. 2024). Observation data is therefore necessary to assess model skill and forecast utility, especially for wind in complex terrain (Jin et al. 2024; T. Nguyen et al. 2023; Ramavajjala and Mitra 2023). The WeatherReal benchmark dataset addresses the need for observation-based evaluation by compiling and performing quality control on in-situ measurements. The publicly available version of this dataset, WeatherReal-ISD, includes only approximately 13,000 stations globally and does not provide near-surface wind gust observations (Jin et al. 2024). The United States and Europe, which have the highest density of sensor networks, provide the opportunity to construct more extensive regional observation benchmark datasets (Chen and Wang 2026). Leveraging a larger subset of sensor networks available across CONUS, with wind gust observations, allows for the creation of a more representative dataset that better supports the development and evaluation of regional weather forecasts.

While observational data is essential for modeling and decision-making, fragmented observation networks and burdensome quality control have limited the use of available data. Errors can be introduced into the data record from faulty instruments, data recording, transmission, environmental changes, and errors in processing algorithms; therefore, quality control procedures must be applied to ensure data reliability (Petrić et al. 2026; WMO 2010). Quality assurance methods range from simple plausibility checks to complex neighborhood checks (i.e., spatial consistency) that verify measurements against nearby stations. The WeatherReal dataset and standard MADIS quality control levels both employ spatial checks. While valuable for detecting sensor failures, we find that these checks can erroneously reject valid high-wind-speed observations in sparse networks and complex terrain, where wind is highly variable over short distances (see the Section on Technical Validation, Cook 2023). Additionally, MADIS quality control erroneously retains implausible isolated spikes (see the Section on Technical Validation). These deficiencies highlight the need for improved quality control methods to preserve valid high-wind-speed observations while rejecting anomalous measurements. Here, we develop a quality control approach that can differentiate sensor errors from localized extreme wind events common in complex terrain.

To address the gap of a spatially dense, standardized, well-validated observational dataset, we develop the Wind IN-situ Data Benchmark (WIND-Bench) dataset for 2021-2025 near-surface meteorological observations across CONUS. The variables temperature, wind speed, wind direction, wind gust, and relative humidity are included in WIND-Bench because they are used in combination in fire weather decision-making. However, wind is the primary focus of WIND-Bench given the described significant challenges in forecasting and quality control. While developed for high wind events, the upper threshold is 50 m $s^{-1}$ making the dataset likely unsuitable for hurricanes and tornadoes. WIND-Bench integrates MADIS METAR and Mesonet, which aggregates multiple networks, and applies wind-specific quality control designed to retain valid high-wind speed observations. Rather than relying on spatial consistency checks, we verify observations against the National Oceanic and Atmospheric Administration (NOAA) High-Resolution Rapid Refresh (HRRR) model forecast hour 2 (f02) to detect sensor malfunctions while preserving localized extremes. This approach provides a spatially consistent approach to detecting failures rather than relying on an unevenly distributed sensor network. Our approach verifies observations using the time series of error between observations and the reference forecast dataset, which remains within identified bounds during normal sensor operation. The resulting WIND-Bench dataset serves as a resource for training regional ML models and for standardized evaluation and intercomparison of NWP and ML weather forecasts, facilitating accelerated development and deployment of advanced near surface weather forecasts.

# Materials and Methods

WIND-Bench compiles in-situ observation measurements from the MADIS METAR and Mesonet meteorological observation networks and applies a series of quality control algorithms to the stations in each of the 9 National Centers for Environmental Information (NCEI) US Climate Regions (Karl and Koss 1984). Between 2021 and 2025, there were 42,348 MADIS stations operating in CONUS, with as many as 32,520 stations operating in any given year (Figure 1). Station density is highest in coastal regions and near population centers, with the West climate region containing the most stations (Figure 1, Table 1). Table 1 summarizes for each network and climate region the number of stations that report the variables temperature, wind speed, wind direction, wind gust, and relative humidity (Table 1). Temperature is the most widely recorded variable (Table 1).

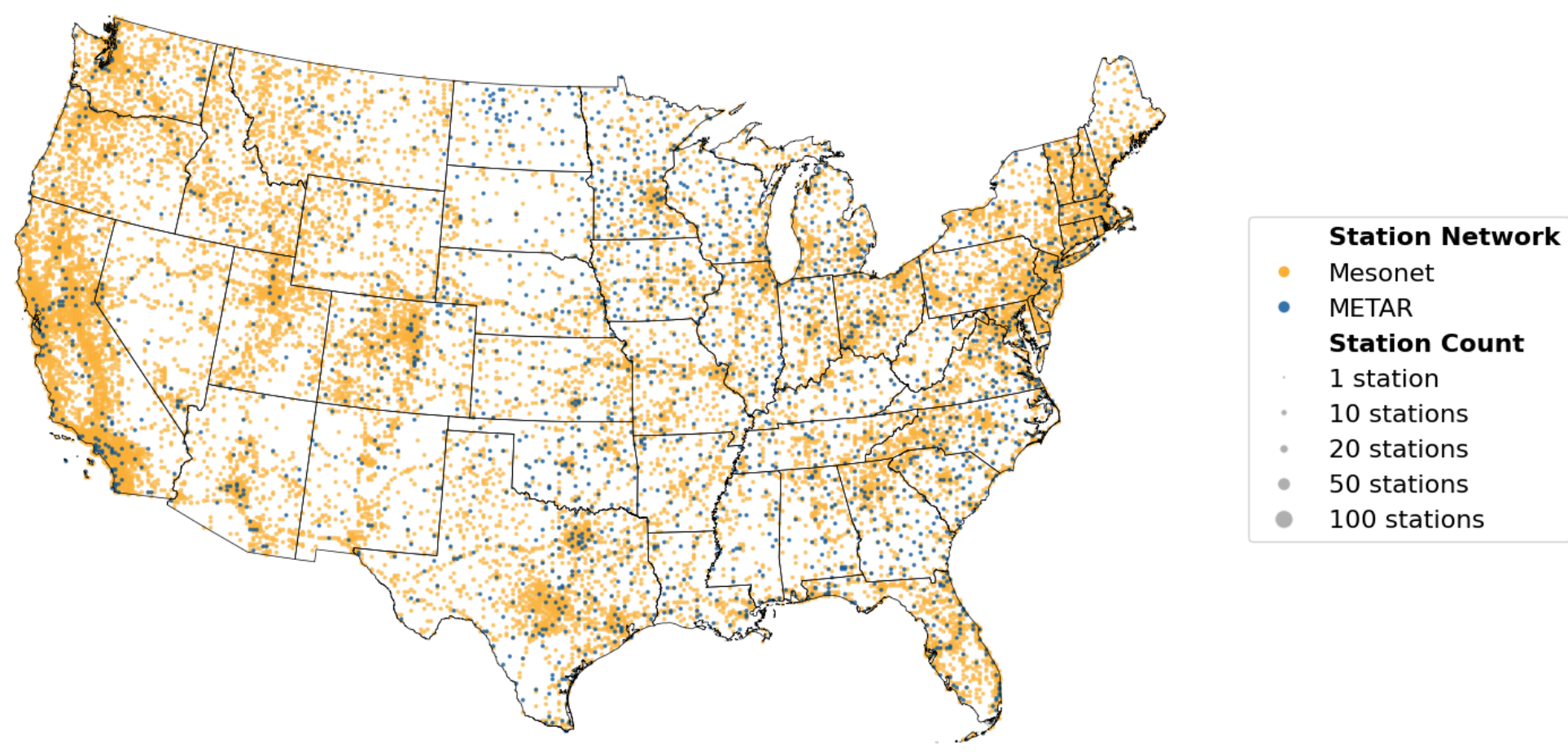

**Figure 1. Map of MADIS observation stations locations and density of MADIS Mesonet and METAR meteorological observation towers. The marker size indicates the density of stations.**

**Table 1. Number of observation stations by climate region and observed variable.**

| | Network | *n Temperature* | *n Wind Speed* | *n Wind Direction* | *n Wind Gust* | *n Relative Humidity* |
|---|---|---|---|---|---|---|
| **Northeast** | Mesonet | 3656 | 2277 | 2246 | 1662 | 3472 |
| | METAR | 206 | 145 | 145 | 0 | 206 |
| **Upper Midwest** | Mesonet | 2087 | 1622 | 1597 | 1521 | 1965 |
| | METAR | 300 | 255 | 254 | 182 | 300 |
| **Ohio Valley** | Mesonet | 2956 | 1869 | 1839 | 1441 | 2683 |
| | METAR | 267 | 200 | 200 | 0 | 267 |
| **Southeast** | Mesonet | 4049 | 2345 | 2317 | 1652 | 4010 |
| | METAR | 377 | 315 | 314 | 0 | 377 |
| **N. Rockies &** | Mesonet | 2267 | 1733 | 1707 | 1518 | 1741 |
| **Plains** | METAR | 190 | 190 | 190 | 116 | 190 |
| **South** | Mesonet | 3416 | 1946 | 1900 | 1729 | 2698 |
| | METAR | 383 | 263 | 262 | 96 | 383 |
| **Southwest** | Mesonet | 4638 | 3808 | 3782 | 3530 | 3891 |
| | METAR | 161 | 161 | 161 | 53 | 161 |
| **Northwest** | Mesonet | 3453 | 2233 | 2205 | 2102 | 2909 |
| | METAR | 97 | 91 | 91 | 11 | 97 |
| **West** | Mesonet | 8310 | 7363 | 7327 | 6851 | 7683 |
| | METAR | 153 | 154 | 154 | 12 | 154 |

# Data Download and Processing

## In-situ Observation Data

We obtained meteorological observation data from NOAA's MADIS network, which compiles observations from sensor networks operated by various organizations (Miller et al. 2007). We used observations from the METAR network, which includes the Automated Surface Observing Systems (ASOS) and Automated Weather Observing Systems (AWOS), and the Mesonet network of networks, which includes the Remote Automatic Weather Stations (RAWS) network and various state, private, and federal networks (NOAA 2017, Table S1). Data was accessed through the public MADIS archive (https://madis-data.ncep.noaa.gov/madisPublic1/data/archive). The observational variables included in the dataset are hourly near-surface wind speed, wind direction, wind gust, temperature, and relative humidity. All measurements are recorded between 1.5 and 10 m of the surface, with height dependent on network and variable with available details in the supplemental material (Table S1).

We downloaded MADIS data without quality control flags to retain the maximum number of observations for subsequent quality control. Each station's time series was resampled to a regular hourly time step, and linear interpolation of up to two consecutive missing time steps was performed to increase temporal completeness while avoiding interpolation over large data gaps. For records where relative humidity was not directly available, it was derived from the near-surface air temperature and dewpoint using the Tetens equation (Tetens 1930).

We chose to omit precipitation from this dataset because more complete precipitation records are available from the NOAA National Weather Service Cooperative Observer Program (COOP), which has operated precipitation gauges since 1895, and currently maintains approximately 8,000 active gauges (Lawrimore et al. 2020). As of 2019, approximately 2,000 of these stations were equipped with Fischer & Porter digital gauges (Lawrimore et al. 2020). Standardized digital sensor technology, precipitation-specific quality control, a long data record, and network density make COOP a superior benchmark for verifying modeled precipitation products (Fang et al. 2022).

The final hourly observations can be treated as instantaneous measurements, although temporal averaging occurs near the top of the hour for select variables and providers (Table S1). As an example of the height and temporal averaging variability present across the networks, RAWS stations report wind speed and direction as 10-minute averages, typically measured at 6.1 m, while air temperature is an instantaneous reading and relative humidity is a 10-minute average, both measured between 1.2 – 2.4 m (Gallagher et al. 2022; NWCG 2019). RAWS defines a wind gust as the maximum hourly wind speed from a minimum of 720 samples (NWCG 2019). Other Mesonet providers use different sensor standards and data-averaging methods; for example, the Oklahoma Mesonet reports all hourly data as 5-minute averages and measures temperature and relative humidity at 1.5 m and wind at 10 m (Oklahoma Mesonet, 2026). In contrast, ASOS stations record wind speed as a 2-minute average of the 5-second average measured between 8 and 10 m and report wind gusts when that average reaches 4.6 m $s^{-1}$ and is exceeded by the 5-second average by 2.6 m $s^{-1}$ (Vickie L. Nadolski 1998). Further details on network-specific station and data-collection specifications are provided in Table S1, where available. The inconsistencies in sensor heights and averaging procedures across sensor networks introduce uncertainty both within the observational dataset and when comparing values against forecasts (Collins et al. 2024b; Fovell and Capps 2024). Sensor height normalization is not possible because sensor height information is neither standardized nor available from all providers (Table S1). Despite these discrepancies, we believe there is still significant value in the extended spatial coverage that combining data across networks provides.

### Forecast Data

Hourly forecasts from NOAA HRRR model Version 4 (HRRRv4) were obtained using the Herbie weather forecast data software package (Blaylock 2025). HRRRv4 provides forecast data at 3 km spatial resolution, issued hourly out to 18 hours ahead and every 6 hours out to 48 hours ahead (Dowell et al. 2022). The 2-hour lead time was selected to provide a reanalysis-like dataset while avoiding the model's initial spin-up period needed to realistically represent atmospheric processes (James et al. 2018). The 2 m temperature, 10 m wind speed, and 2 m relative humidity variables were used for comparisons with in-situ data during quality control.

## Quality Control Methods

A series of quality control checks was applied to each variable to identify erroneous measurements and ensure the reliability of the observational dataset. Table 2 summarizes all quality-control algorithms and thresholds implemented for each variable. Checks were applied to hourly data with the full quality control pipeline run separately for each year and climate zone.

Filters were applied in sequential stages: an initial set was applied to the raw data, and subsequent filters were applied to the refined dataset. This approach has several benefits, including allowing the isolated spike filter to be applied twice catching spikes of two observations and enabling the mean absolute error filter, which relies on calculated statistics, to be applied after extreme outliers were removed. The stage or stages at which each filter is applied are indicated in Table 2.

**Table 2. Summary of quality control algorithms applied to each variable. The number in parentheses indicates the stage or stages during which the quality control algorithm was applied.**

| | WS (m $s^{-1}$) | T (°C) | WD (°) | WG (m $s^{-1}$) | RH (%) |
|---|---|---|---|---|---|
| **Linear Gaps** | ✓(1) | ✓(1) | ✓(1) | ✓(1) | ✓(1) |
| N Observations | 4 | 4 | 4 | 4 | 4 |
| **Constant Value** | ✓(1) | ✓(1) | ✓(1) | ✓(1) | ✓(1) |
| **Value Range** | ✓(1) | ✓(1) | ✓(1) | ✓(1) | ✓(1) |
| Lower Limit | 0 | -60 | 0 | 0 | 0 |
| Upper Limit | 50 | 60 | 360 | 66 | 100 |
| **Spikes** | ✓(1,2) | ✓(1,2) | | ✓(1,2) | |
| Difference Threshold | 20 | 15 | | 20 | |
| Extreme Percentile | 99 | 98 | | 99 | |
| **Spike Groups** | ✓(1,2) | | | ✓(1,2) | |
| N in group | 3 | | | 3 | |
| Difference Threshold | 20 | | | 20 | |
| **Percentile** | ✓(1) | | | ✓ (1) | |
| $10^{th}$ p threshold | Med. + 6*MAD | | | Med. + 8*MAD | |

| | WS (m s$^{-1}$) | T (°C) | WD (°) | WG (m s$^{-1}$) | RH (%) |
|---|---|---|---|---|---|
| 95$^{th}$ p threshold | 1 | | | 2 | |
| **Mean Absolute Error** | ✓ (2) | ✓(2) | | ✓(2) | ✓(2) |
| Threshold | 8 | 8.5 | | 13 | 45 |
| **Change Point** | ✓(1) | ✓(1) | | ✓(1) | ✓(1) |
| 50$^{th}$ p threshold | 8 | 10 | | 8 | 30 |
| Length (hr) | 168 | 24 | | 168 | 168 |
| **Point Absolute Error** | ✓(1) | ✓(1) | | ✓(1) | ✓(1) |
| Threshold | 45 | 8.5 | | 43 | 56 |
| **Completeness** | ✓(1,3) | ✓(1,3) | ✓(1,2) | ✓(1,3) | ✓(1,3) |
| Completeness (%) | 8, 50 | 8, 50 | 8, 50 | 8, 50 | 8, 50 |
| Window (hr) | 8760, 24 | 8760, 24 | 8760, 24 | 8760, 24 | 8760, 24 |

## Station Quality Control

Stations with fewer than one month of observations throughout the year were removed because this yielded insufficient data to reliably compute statistics and apply time-series quality-control algorithms.

## Time Series Quality Control

### *Linear Gaps*

Some datasets may have had gaps filled by their providers using constant values or linear interpolation over time. We wanted to identify these periods and flag them, as they are not representing true observations. We flag periods where there are four or more consecutive observations that change linearly over time. These segments are identified using a second-derivative threshold to detect unrealistically low curvature (Table 2). While periods of meteorologic stasis and linear change are possible, several consecutive linear hours are unlikely to occur naturally. .

### *Constant Values*

Stations reporting a constant value may indicate a stuck or faulty sensor. If the minimum and maximum observation across the time series were equal, the entire station time series was flagged and masked. This filter identifies stations missed by the linear gaps filter, because it evaluates all valid observations rather than consecutive measurements.

### *Spikes & Spike Groups*

Isolated extreme spikes or spike groups of multiple extreme values appear in some time series data. An observation was flagged as an isolated spike if it both exceeded a regional extreme percentile value and had an absolute difference between the extreme value and one or both immediate neighbors exceeding the variable-specific threshold (Table 2). Percentile thresholds were determined through visual inspection of time series in the Southwest region, where the 99$^{th}$ percentile 10.9 m s$^{-1}$ for wind speed, 18.3m s$^{-1}$ for wind gusts, and 33.4°C for temperature. Spike groups were identified as up to 3 consecutive observations that exceeded the regional variable-specific extreme-value threshold and were bounded by large jumps both before and after the spike group (Table 2).

### *Value Range*

Value range checks were applied to identify and mask measurements that fall outside physically plausible limits based on NOAA MADIS validity checks and NOAA National Center for Environmental Information records (Table 2) (NCEI 2026; NOAA 2017). Observations below the lower threshold and above the upper threshold were flagged and masked.

### *Persistence of Regionally High or Low Values*

Some stations exhibited persistently anomalously high or low wind speed and wind gust measurements for an extended period. This filter is designed to catch stations with sensor drift from degraded bearings or poor anemometer calibration (Azorin-Molina et al. 2018; Paulsen et al. 2007). Stations with anomalously high wind speeds were identified if their $10^{th}$ percentile wind speed value was 6 standardized median absolute deviations (MADs) from the regional median $10^{th}$ percentile. Similarly, stations with anomalously high wind gusts were identified using a threshold of 8 MADs from the regional median $10^{th}$ percentile. The median was chosen to characterize the central tendency of the region's $10^{th}$ percentile distribution while being robust to outliers. Additionally, stations where the $95^{th}$ percentile wind speed was below 1 m $s^{-1}$ and wind gusts below 1 m $s^{-1}$ were removed. Although anomalously low wind-speed stations may result from extreme sheltering by terrain or the built environment rather than an anemometer malfunction, these stations were masked because the data did not represent the surrounding meteorological conditions.

### *Rolling Completeness*

Sensor malfunctions or data transmission issues can cause frequent periods of missing observations, which limit the efficacy of quality control algorithms. To identify periods of missing data, we applied a rolling completeness check twice during the quality control process. First, to ensure that a minimum of a month of data was present for the station across the year, and second, to identify periods with more than 50% of observations missing within a 24-hour rolling window. If windows do not meet that threshold, all data within that 24-hour period was flagged and masked. We acknowledge this filter may introduce selection bias if sensors are less reliable during extreme events. Our extensive evaluation of time series indicates that intermittent completeness is more often coupled with other sensor malfunctions than with extreme winds, suggesting this filter primarily removes poor data quality rather than extreme events.

## Verification with HRRR

Comparing observation data to a reference dataset, such as reanalysis or forecast data, helps detect changes in sensor equipment and data processing in meteorological time series (K. N. Nguyen et al. 2021; Vejen et al. 2002). Although all reference datasets contain some bias, the bias is generally small and follows relatively consistent patterns, making comparison valuable for assessing whether measurements are physically reasonable and for detecting dramatic shifts that could indicate sensor malfunction. We use the HRRRv4 f02 hourly forecasts in our observation quality control methods to assess when observations and time periods deviate from the forecast by amounts exceeding the expected bias magnitude. Each observation station is paired with the nearest HRRR grid cell using a nearest-neighbors algorithm, and observation times are matched to the corresponding forecast's valid time. Observed surface temperature, wind speed, wind gusts, and relative humidity were quality controlled using HRRR 2m temperature, 10 m wind speed, 10 m gust potential, and 2 m relative humidity, respectively.

Surface winds in the HRRR forecast are represented at 10 m above ground level, while anemometer heights within station networks vary between 2 and 10 m. Because wind speeds are slower near the surface, this discrepancy in measurement height may introduce additional bias between observations and forecasts at some stations. HRRR-forecasted gust potential is intended to represent the upper bound of possible gusts and is therefore used to verify near-surface wind gust observations; however, it tends to underpredict observed gusts, potentially due to the model's boundary layer physics or how the properties of the land-surface impact the gust (Collins et al. 2024b; Glasser 2025; Collins et al. 2024a).

*Mean Absolute Error*

Stations with annual mean absolute errors (MAE) exceeding thresholds of 8 m $s^{-1}$ for wind speed, 8°C for temperature, 13 m $s^{-1}$ for wind gusts, and 45% for relative humidity were flagged and masked (Table 2). The MAE is calculated as the mean of the absolute differences between observations and the HRRR time series across the annual time series at a station.

Thresholds were defined based on the distribution of MAE values across all stations. Wind speed and wind gust thresholds correspond to the 99.5$^{th}$ percentile of the error distribution, while the temperature and relative humidity threshold corresponds to the 98.5$^{th}$ percentile due to a long tail observed in the temperature error distribution. While the MAE varies across stations with location, these thresholds are set to capture stations with uncalibrated sensors and sensor drift..

*Hourly Absolute Error*

The difference between hourly observations and HRRR was computed for wind speed, wind gust, temperature, and relative humidity. Observations with an absolute error exceeding 45 m $s^{-1}$, 43 m $s^{-1}$, 8.5°C, and 56.0%, respectively, were flagged and masked. Thresholds for temperature and relative humidity were defined as the 99$^{th}$ percentile of the forecast error across all stations. A percentile approach was not chosen for wind because wind speed and gust error distributions are highly non-Gaussian, with higher wind speeds associated with greater forecast error. Wind thresholds were defined based on Collins et al. (2024), who evaluated the error between observations and HRRR forecast hour 6 across the Colorado-Wyoming region (Collins et al. 2024a). The goal of this filter was to balance removing points associated with sensor error, while retaining high wind speeds, which are known to have large forecast errors.

*Error Change Point Detection*

A change point detection algorithm from the Ruptures Python package is used to detect abrupt changes in the HRRR observation error time series (Truong et al. 2020). Since the bias in HRRR forecasts for an individual grid cell is generally consistent within certain bounds, abrupt changes in the bias time series were used to identify potential sensor malfunctions. We used a least-absolute-deviation cost function to detect regions with variable-dependent minimum window sizes to detect shifts in sensor behavior (Table 2). Extensive trial and error was used to define thresholds and segment windows rather than a formal optimization method. For Temperature observations, segments between change points were flagged and masked when the median error exceeded 10 ºC for 24 hours or longer (Figure 2a, Table 2). For wind speed and wind gust observations, segments between change points were flagged and masked when the median error exceeded 8 m $s^{-1}$ for 168 hours or longer (Figure 2b, Table 2).

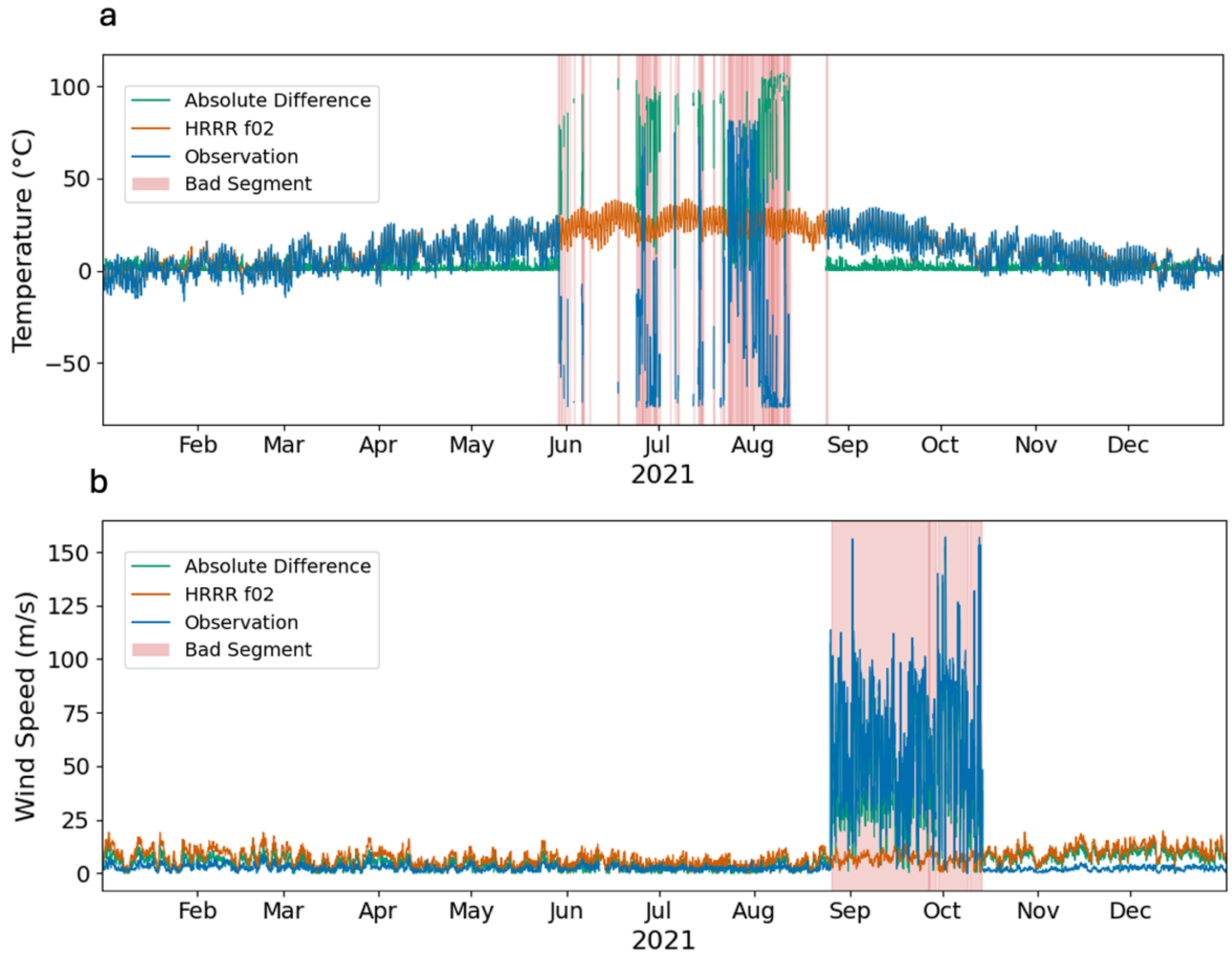

**Figure 2. Change point detection quality control filter examples time series (a) temperature and (b) wind speed.**

# Data Records

WIND-Bench is publicly available from the Open Energy Data Initiative (OEDI) (see Section on Data Availability). This dataset provides quality-controlled hourly meteorological observations for near-surface wind speed, wind direction, temperature, relative humidity, and wind gusts at stations across CONUS for 2021-2025.

The dataset is available as annual NetCDF files with float32 precision. Spatial coordinates are in the WGS 84 reference system, and time is recorded in Coordinated Universal Time (UTC). Station network, data provider, station type, station ID, and station name are included as metadata within the NetCDF file where available.

# Technical Validation

To demonstrate the quality of WIND-Bench, we evaluated variable distributions, the goodness of fit of observations to the HRRR reference dataset, comparison to MADIS quality control levels for high wind speeds, and goodness-of-fit ot HRRR during an extreme wind and temperature event.

## Observation Evaluation

Figure 3 shows the data distributions for five variables across the nine climate regions. For calm winds of 0 m $s^{-1}$, wind direction is recorded as 0° or 180° depending on the sensor network. These values were excluded from Figure 3c to better represent the wind direction distribution. The distributions and regional variability in the dataset match expected trends. For example, the highest median temperature, 20.7°C, is in the South, while the lowest, 7.9°C, is in the Northern Rockies and Plains Climate Region (Figure 3a). The wind speed and wind gust distributions are skewed, with long tails and fewer observations at high wind speeds (Figure 3b, 3d). Additionally, the lowest median relative humidity, 43.9%, is observed in the arid Southwest, while the highest, 75.9%, is observed in the humid Southeast (Figure 3e).

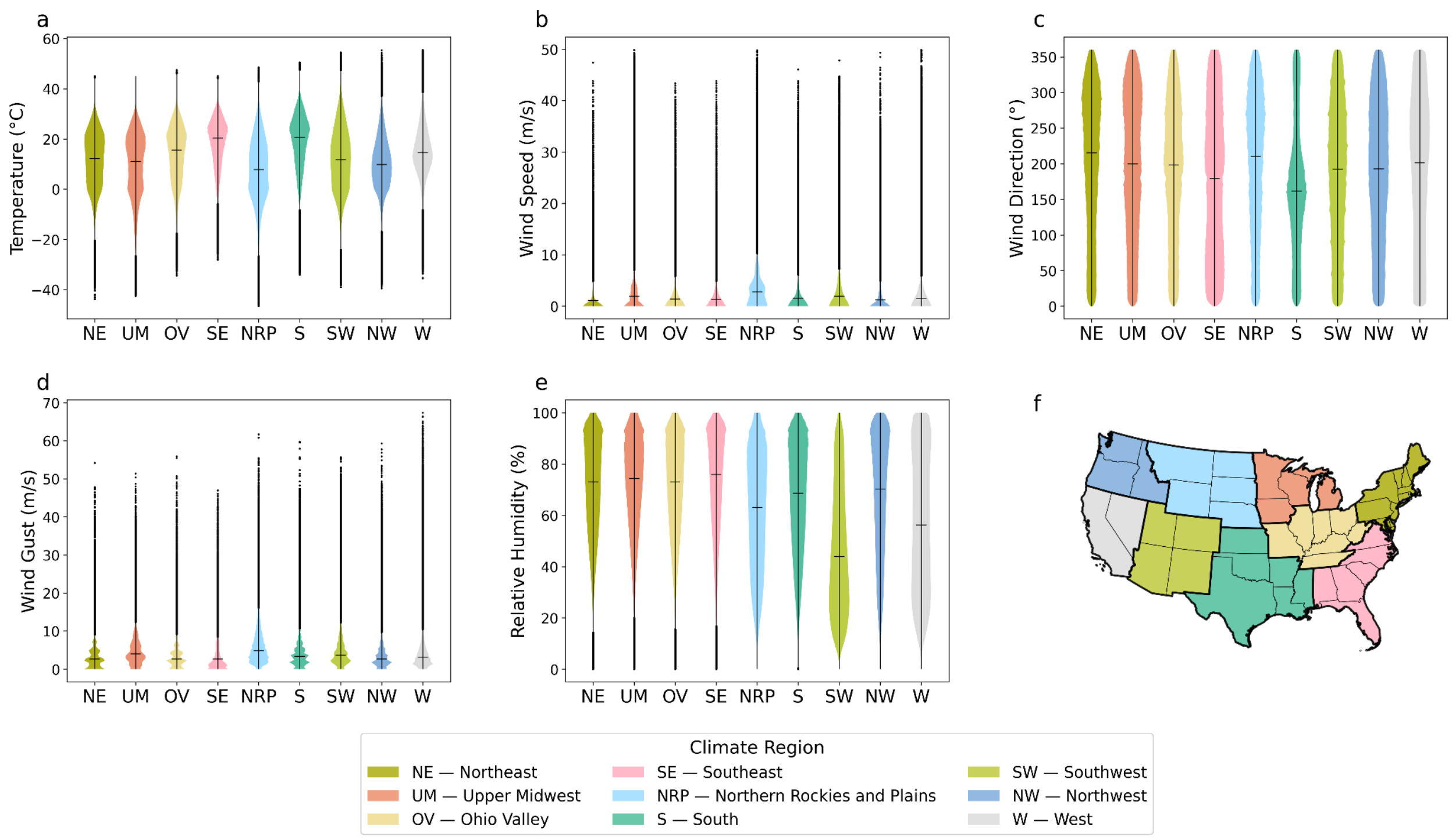

**Figure 3. Distributions of regional meteorological variables for 2021-2025. (a) Temperature (b) wind speed (c) wind direction, where wind is greater than 0 m $s^{-1}$ (d) wind gust (e) relative humidity, and (f) map of climate regions. The horizontal black line indicates the median; the shaded region represents the interquartile range (IQR); the vertical black line extends to ±1.5 × the IQR; and the black dots represent outliers.**

Outliers visible in Figure 3 correspond with documented extreme weather events. For example, the maximum temperature of 56°C recorded in the Northwest region occurred on 29 June 2021 at station UR058 operated by Union Pacific Railroad with data provided by MesoWest, during the peak of a record-breaking regional heatwave (MesoWest 2026; Werth et al. 2011; White et al. 2023). For context, the highest temperature reported elsewhere in Oregon during this event was 48.3°C observed at both Moody Farms and Pelton Dam, 102 and 123 km southwest of station UR058, respectively (Vescio and Bair 2022). Additionally, the Dalles Municipal Airport METAR station, located 50 km west of UR058 along the Columbia River, recorded a temperature of 47.8°C (Clarkson et al. 2023). Although temperatures reported at UR058 exceed other recorded maximum temperatures, these coincide spatially and temporally with this extreme heat event. These extreme values may represent localized conditions or poor sensor placement rather than instrument error and should not be arbitrarily removed.

## Verification Against MADIS Quality Control

WIND-Bench was evaluated against data with MADIS Level 3 quality-controlled data for a high wind event in Boulder County, Colorado, for stations across the entire state of Colorado in 2021, and for two stations with extreme high wind observations exceeding 50 m $s^{-1}$ in the Gulf Coast. Level 3 is the strictest MADIS quality control level, utilizing a number of statistical quality control filters in addition to a spatial consistency check.

On 30 December 2021, Boulder County experienced an extreme downsloping windstorm leading to the catastrophic Marshall Fire (Benjamin et al. 2023). At 14:00 MST during the high wind event, we find that the MADIS level 3 algorithm rejects 33 of 66 wind speed observations, including observations from all 4 METAR stations in the region. Meanwhile, WIND-Bench rejected only 4 out of 66 observations. After MADIS quality control, the mean wind speed measured in Boulder County was 5.94 m $s^{-1}$, with a maximum of 32.1 m $s^{-1}$. By contrast, WIND-Bench had an average wind speed of 9.8 m $s^{-1}$ and a maximum wind speed of 33.8 m $s^{-1}$. In Figure 4 we see that the MADIS quality control consistently selects low wind speeds while rejecting high wind speeds. This pattern is exemplified by MADIS rejecting a cluster of observations exceeding 20 m $s^{-1}$ at the intersection of State Highways 72 and 93 (39.87, -105.24 in Figure 4b) close to the highest observed wind gusts reported by others (Benjamin et al. 2023). Accordingly, we propose that our quality control approach better preserves high wind speeds during this known extreme wind event (Figure 4).

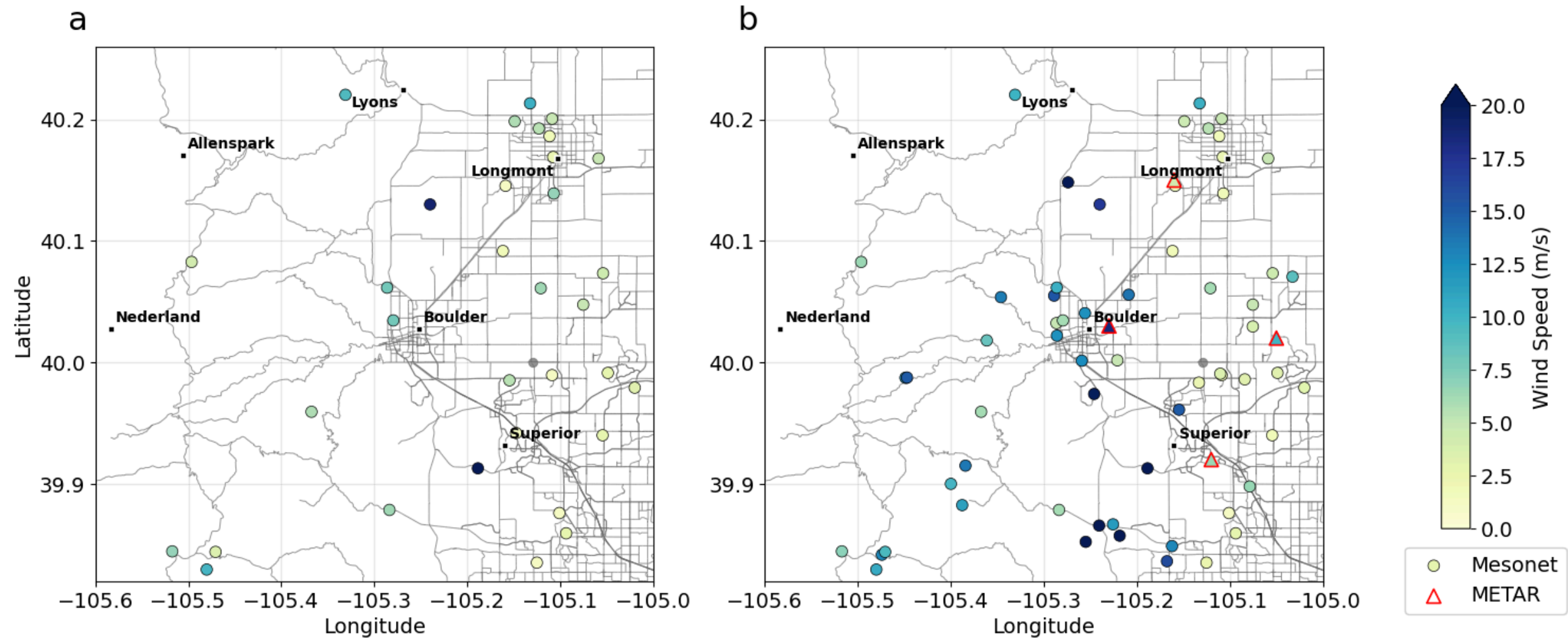

**Figure 4. Observed speed in the Boulder, CO area on December 30th, 2021 at 2:00 PM MST (a) MADIS quality control level 3 (b) WIND-Bench dataset.**

Table 3 summarizes the wind speed observations from all METAR and Mesonet stations across Colorado for 2021 for the two different quality control approaches. It shows the total number of available observations (no quality control), observations retained by MADIS Level 3 quality control, and observations retained in the WIND-Bench dataset. For wind speeds exceeding 10 m $s^{-1}$, WIND-Bench retains 96.7% of observations compared to 52.0% retained by MADIS level 3 quality control (Table 3). Based on these counts, the MADIS quality control appears to be rejecting many real wind observations in the 10-30 m $s^{-1}$ range while preserving erroneous measurements exceeding 50 m $s^{-1}$. This finding is consistent with previous work that showed ASOS real-time quality control to erroneously remove wind gust observations during convective events (Cook 2023). We take a closer look at some of these extreme measurements in the next paragraph.

**Table 3: Number of wind speed observations by 5 m $s^{-1}$ bin for the total available observations, retained after MADIS level 3 quality control, and observations retained in WIND-Bench for the METAR and Mesonet networks for Colorado in 2021.**

| Wind Speed (m $s^{-1}$) | Total | MADIS Level 3 | WIND-Bench |
|---|---|---|---|
| **0-5** | 6005471 | 5379058 | 5021790 |
| **5-10** | 622533 | 571079 | 614285 |
| **10-15** | 64282 | 36076 | 63011 |
| **15-20** | 8162 | 2462 | 7722 |
| **20-25** | 1371 | 130 | 1195 |
| **25-30** | 191 | 8 | 133 |
| **30-35** | 67 | 2 | 22 |
| **35-40** | 31 | 2 | 3 |
| **40-45** | 44 | 11 | 0 |

| 45-50 | 67 | 9 | 0 |
|---|---|---|---|
| 50+ | 312 | 51 | 0 |

To assess whether wind speeds exceeding 50 m $s^{-1}$ in the MADIS level 3 dataset represent physically plausible observations, we evaluated two stations on the Gulf Coast hypothesizing that these observations may be due to extreme events like hurricanes making landfall. However, a station in southern Texas recorded a wind speed of 73.9 m $s^{-1}$ on 22 December 2021 at 14:00 UTC, which is not associated with any recorded hurricane (Figure 5a). Similarly, a station on the Gulf Coast of Florida recorded wind speeds of 108.7 m $s^{-1}$ on 14 March 2021 21:00 UTC, 111.9 m $s^{-1}$ on 27 March 2021 11:00 UTC, and 108.8 m $s^{-1}$ on 24 April 2021 at 19:00 UTC (Figure 5b). These are each isolated high wind observations (e.g., one or two anomalous datum in otherwise quiescent conditions) that cannot be attributed to known high wind events such as hurricanes. The Spike and Point Absolute Error filters applied to WIND-Bench effectively remove these occurrences, providing more accurate quality control than MADIS level 3.

These findings lead us to conclude that the MADIS Level 3 quality control is indeed discarding known high wind events while preserving anomalous extreme measurements that are more likely due to sensor problems rather than real weather.

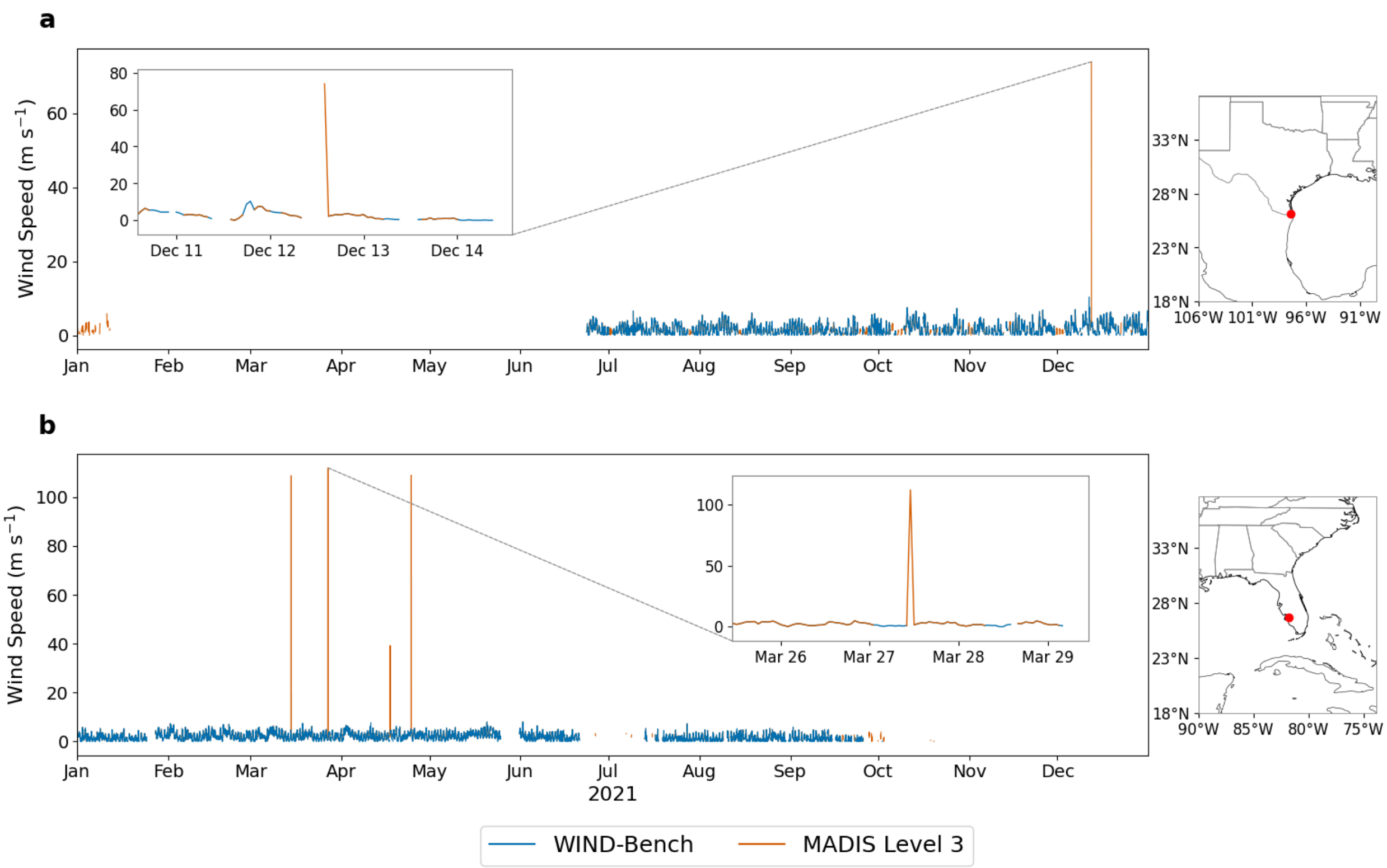

**Figure 5. Wind speed observations for 2021 from WIND-Bench and MADIS level 3 dataset for two stations on the Gulf Coast. Time series of hourly wind speed (m $s^{-1}$) for (a) a station in southern Texas and (b) a station in southwestern Florida. Insets show a four day window surrounding the largest spike from each station. Station location is represented by a red marker on the maps to the right of the timeseries.**

## Verification Against HRRR

We demonstrate the validity of WIND-Bench compared to the HRRR forecasts as a reference. Observation stations were matched to the nearest HRRR grid cell using a nearest-neighbor lookup, and only stations with at least half a year of valid data across the study period were used to calculate goodness-of-fit statistics. This verification is completed across the full value range of variables. Wind direction error calculated as the shortest angular distance between the forecast and observed directions over the range 0° to 180° for observations with wind speed > 0 m $s^{-1}$.

The differences between the datasets, measured by Root Mean Squared Error (RMSE), are heterogeneous across CONUS (Fig. 4). Temperature RMSE is greatest in the southwest for Mesonet stations (2.24°C) and METAR stations (1.81°C) (Table 4, Figure 6a), with the highest values concentrated in mountainous regions. When evaluating all Mesonet stations, wind speed RMSE is higher in the Northeast (3.10 m s-1) (Table 4). However, when evaluating METAR stations, wind speed RMSE is greatest in the Southwest (1.92 m s-1), while RMSE is 1.54 m $s^{-1}$ in the Northeast (Table 4). For temperature, wind direction, and relative humidity, RMSE is highest in mountainous regions, with better-performing forecasts in the Upper Midwest, South, and over the Plains (Table 4, Figure 6a,c,d).

**Table 4. RMSE between WIND-Bench observations and HRRR f02 forecast by climate region. RMSE is computed for each station over 2021-2025, then averaged across stations within each region. Stations with at least 4,300 valid observations (approximately 10% of data across the study period) are used to calculate RMSE, and metrics are reported for METAR and Mesonet, which encompasses all remaining sensor networks.**

| | Network | Temperature (ºC) | Wind Speed (m $s^{-1}$) | Wind Direction (º) | Relative Humidity (%) |
|---|---|---|---|---|---|
| **Northeast** | Mesonet | 1.63 | 3.10 | 64.75 | 11.83 |
| | METAR | 1.52 | 1.54 | 53.10 | 10.10 |
| **Upper Midwest** | Mesonet | 1.61 | 2.62 | 51.84 | 10.43 |
| | METAR | 1.44 | 1.47 | 43.01 | 9.33 |
| **Ohio Valley** | Mesonet | 1.65 | 3.03 | 59.74 | 11.48 |
| | METAR | 1.33 | 1.48 | 48.79 | 9.48 |
| **Southeast** | Mesonet | 1.59 | 2.67 | 63.25 | 11.61 |
| | METAR | 1.41 | 1.50 | 54.09 | 9.72 |
| **N. Rockies & Plains** | Mesonet | 2.19 | 2.35 | 62.28 | 13.24 |
| | METAR | 1.66 | 1.77 | 50.74 | 10.64 |
| **South** | Mesonet | 1.68 | 2.82 | 58.22 | 10.16 |
| | METAR | 1.36 | 1.53 | 48.40 | 8.76 |
| **Southwest** | Mesonet | 2.24 | 2.53 | 72.41 | 13.33 |
| | METAR | 1.81 | 1.92 | 65.25 | 10.90 |
| **Northwest** | Mesonet | 2.08 | 2.26 | 76.07 | 14.14 |

| | Network | Temperature (ºC) | Wind Speed ($m\ s^{-1}$) | Wind Direction (º) | Relative Humidity (%) |
|---|---|---|---|---|---|
| | METAR | 1.67 | 1.71 | 70.59 | 11.56 |
| **West** | Mesonet | 2.13 | 2.17 | 72.20 | 13.84 |
| | METAR | 1.66 | 1.67 | 68.13 | 10.78 |

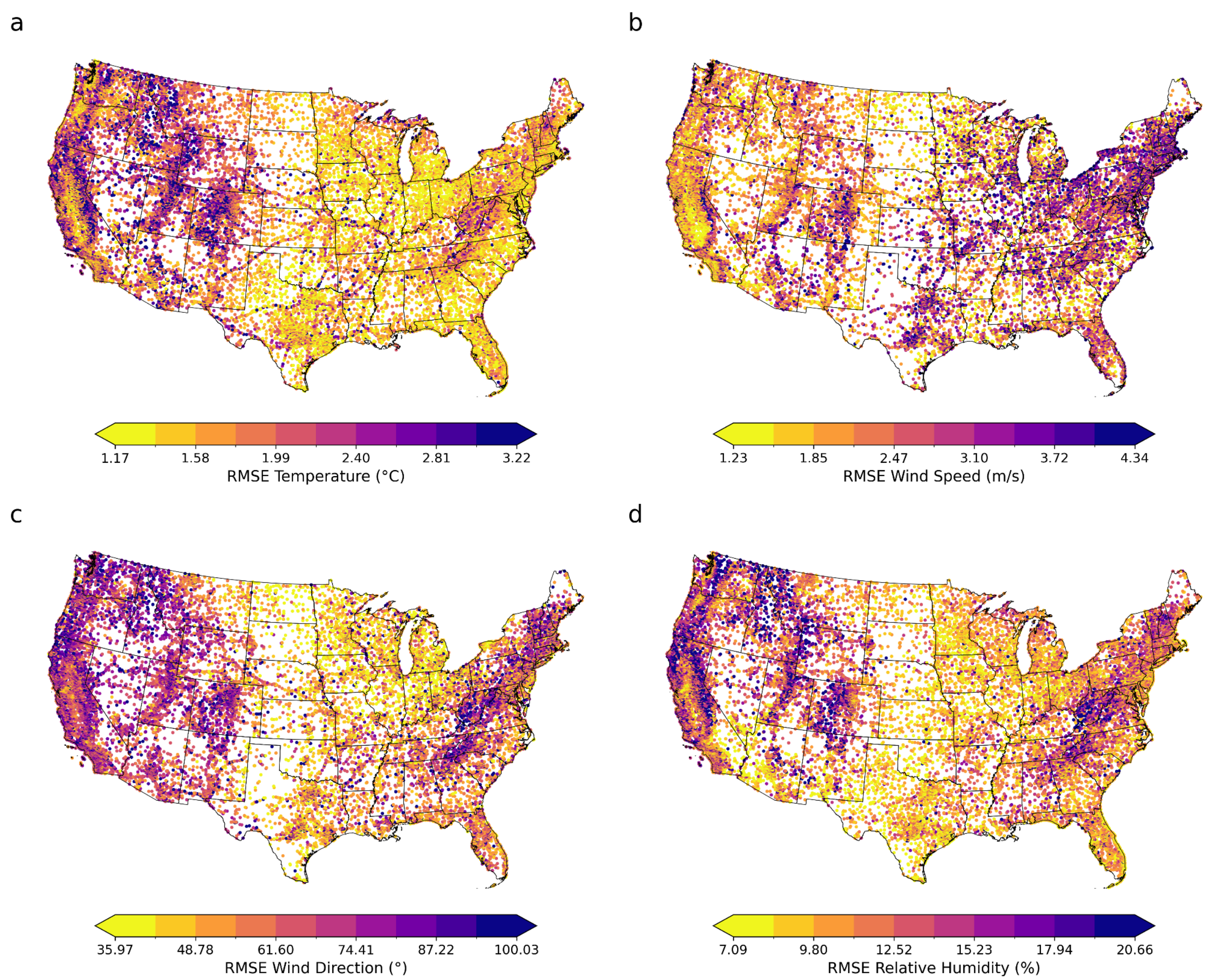

**Figure 6. RMSE between HRRR f02 and WIND-Bench observations for 2021-2025 (a) temperature, (b) wind speed, (c) wind direction, and (d) relative humidity.**

Consistent with previous literature, we find that the HRRR model systematically overestimates wind speed across all regions of CONUS, with the greatest positive bias from Mesonet stations in the Northeast (Table 5, Figure 7b) (James et al. 2022). While the HRRR exhibits a positive bias overall, this is dominated by the overestimation of more prevalent low wind speeds while the model underestimates high wind speeds (Collins et al. 2024a). This positive bias is greater when validated using Mesonet network stations compared to METAR only observations, with the largest discrepancy in the Northeast where Mesonet Bias is of 2.66 $m\ s^{-1}$ while METAR bias is 0.39  $m\ s^{-}$ (Table 5).

The discrepancy is likely explained by station placement bias. Complex terrain and canopy cover both influence HRRR wind speed forecast accuracy (D. Liu et al. 2025). Stations in the METAR network are located at airports, which are typically in flat canopy clearings with low surface roughness. As a result, error statistics based only on METAR stations may underestimate the forecasted wind speed error in regions with more canopy cover, particularly in the Eastern United States. The larger overestimation of wind speed in the Mesonet network may also result from variable anemometer heights, with some wind speeds measured below 10 m. HRRR forecasts wind speed at 10 m, so anemometers mounted lower would record lower wind speeds, which could result in greater average model overestimation.

**Table 5. Error between observations and HRRR f02 forecast by climate region. Mean bias error (MBE) is calculated for temperature, wind speed, and relative humidity and Mean Absolute Error (MAE) for wind direction. Bias error is calculated as HRRR minus observation. Error is computed for each station over 2021-2025, then averaged across stations within each region. Stations with at least 4,300 valid observations (approximately 10% of the data across the study period) are used to calculate error, and metrics are reported for METAR and Mesonet, which encompass all remaining sensor networks.**

| | Network | Temperature (ºC) | Wind Speed (m $s^{-1}$) | Wind Direction (º) | Relative Humidity (%) |
|---|---|---|---|---|---|
| **Northeast** | Mesonet | -0.05 | 2.66 | 50.35 | -4.78 |
| | METAR | -0.22 | 0.39 | 35.56 | -0.30 |
| **Upper Midwest** | Mesonet | 0.00 | 2.06 | 37.60 | -3.64 |
| | METAR | -0.11 | 0.48 | 26.29 | -1.48 |
| **Ohio Valley** | Mesonet | 0.03 | 2.60 | 45.52 | -4.86 |
| | METAR | -0.06 | 0.58 | 31.21 | -1.82 |
| **Southeast** | Mesonet | -0.03 | 2.20 | 48.19 | -4.95 |
| | METAR | -0.04 | 0.59 | 35.92 | -2.29 |
| **N. Rockies & Plains** | Mesonet | 0.16 | 1.02 | 46.65 | -6.57 |
| | METAR | 0.10 | -0.20 | 34.17 | -4.34 |
| **South** | Mesonet | -0.03 | 2.32 | 43.41 | -3.31 |
| | METAR | -0.07 | 0.53 | 31.03 | -1.64 |
| **Southwest** | Mesonet | 0.12 | 1.45 | 56.20 | -6.87 |
| | METAR | 0.21 | -0.10 | 47.63 | -5.11 |
| **Northwest** | Mesonet | 0.10 | 1.50 | 60.92 | -6.47 |
| | METAR | 0.09 | 0.02 | 53.21 | -3.57 |
| **West** | Mesonet | 0.11 | 1.23 | 56.46 | -6.16 |
| | METAR | -0.02 | -0.10 | 50.45 | -2.38 |

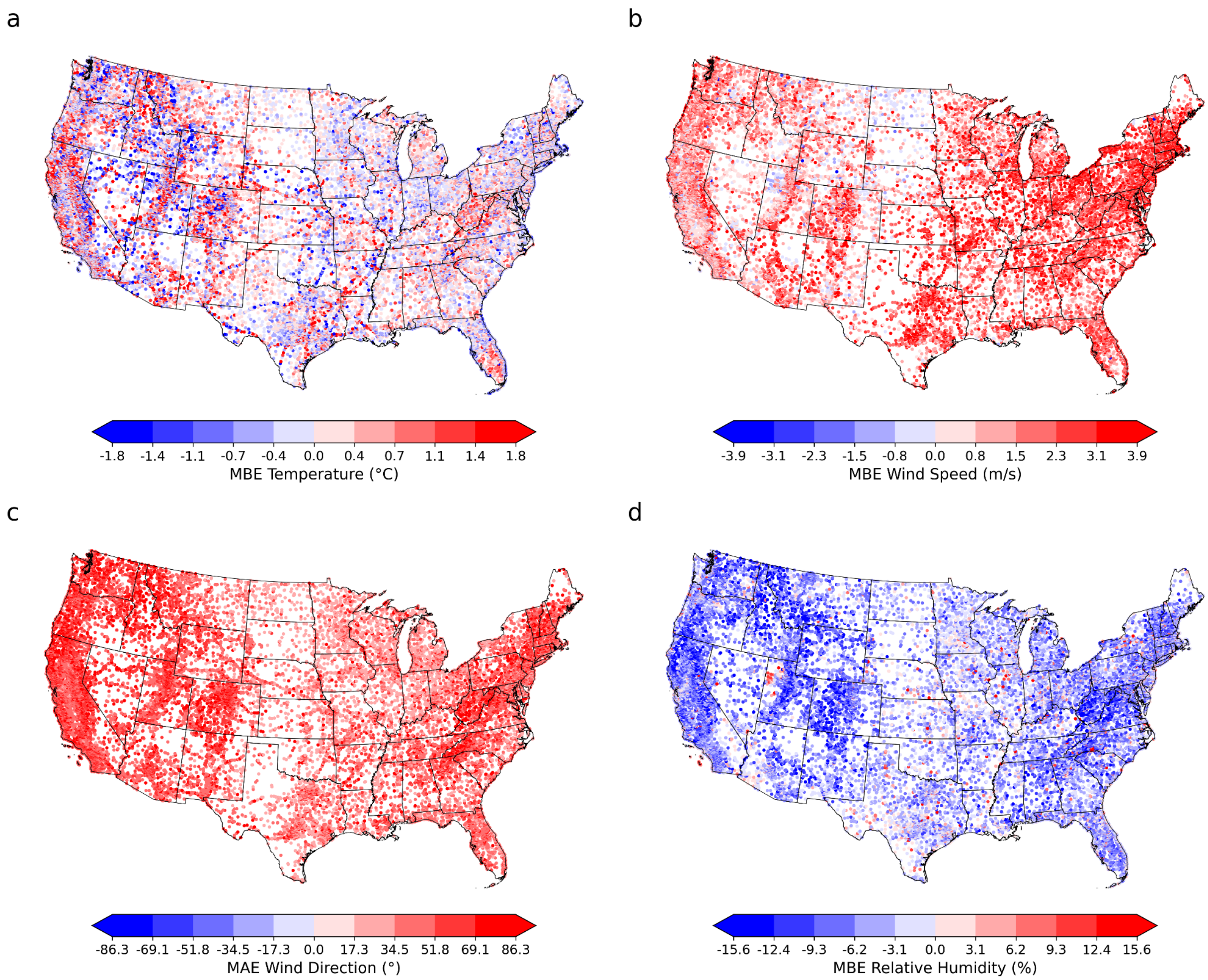

**Figure 7. Error between HRRR f02 and WIND-Bench observations for 2021-2025. (a) MBE Temperature (b) MBE wind speed (c) MAE wind direction and (d) MBE relative humidity.**

## Verification of a Cold Front Event

To evaluate WIND-Bench's ability to represent synoptic-scale events in near-surface conditions, we examined the spatiotemporal evolution of temperature and wind speed during the December 2022 Arctic outbreak. This cold-front event was chosen because the temperature and wind shifts associated with cold fronts are generally more dramatic than those of other types of surface fronts, and these shifts can impact fire ignition and spread (Schultz 2005; Ma et al. 2010). Elevated wind speeds developed and propagated across the Great Plains on 22 December 2022 in both the observations and the HRRR forecasts (Fig. 8). This side-by-side comparison between MADIS quality-controlled observations and forecasted values indicates that the observation density and applied quality control are sufficient to resolve large-scale synoptic patterns. For this event, the HRR model predicts wind speeds along the storm front in the south, while underpredicting wind speeds in the Midwest (Figure 8).

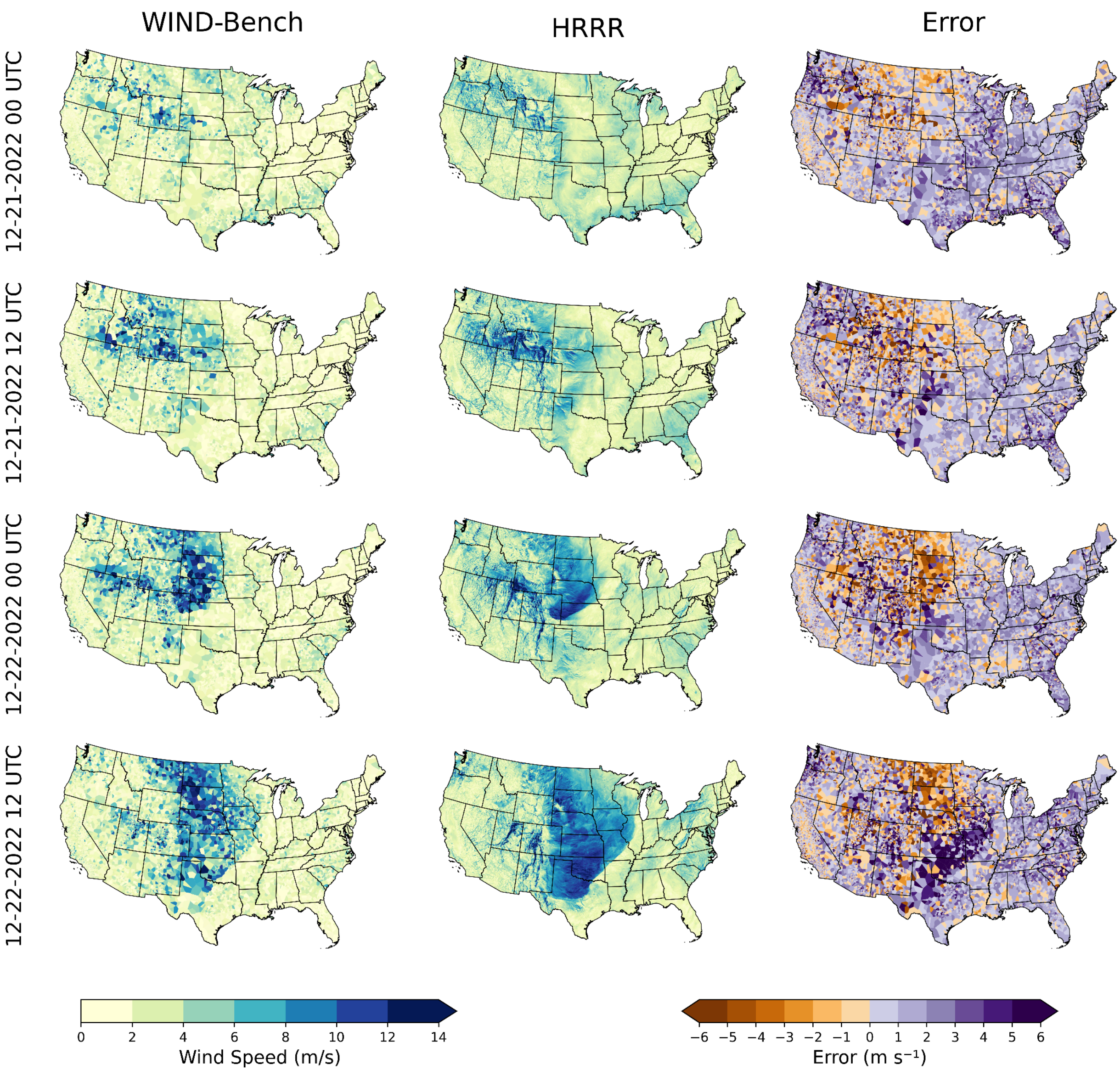

**Figure 8. Surface wind speeds across CONUS on December 21-22, 2022. Left column: WIND-Bench observed wind speeds, shown as a Voronoi diagram. Middle column: forecasted 10 m wind speeds from the HRRR model. Right column: wind speed error, calculated as HRRR-forecasted minus WIND-Bench, shown as a Voronoi diagram. Rows show valid times 12 hours apart beginning at 12-21-2022 00 UTC.**

WIND-Bench captures the southward advance of cold temperatures forecasted in the HRRR model (Fig. 9). The error between the observations and forecasted temperatures varies with forecast hour (Figure 9). At the 00 UTC time steps HRRR underpredicts temperature more, underpredicting temperature across the West and Southwest and in the Northeast (Figure 9). Meanwhile at the 12 UTC time step, HRRR overpredicts temperature in the Northeast and underpredicts temperature the most along the leading edge of the cold front (Figure 9).

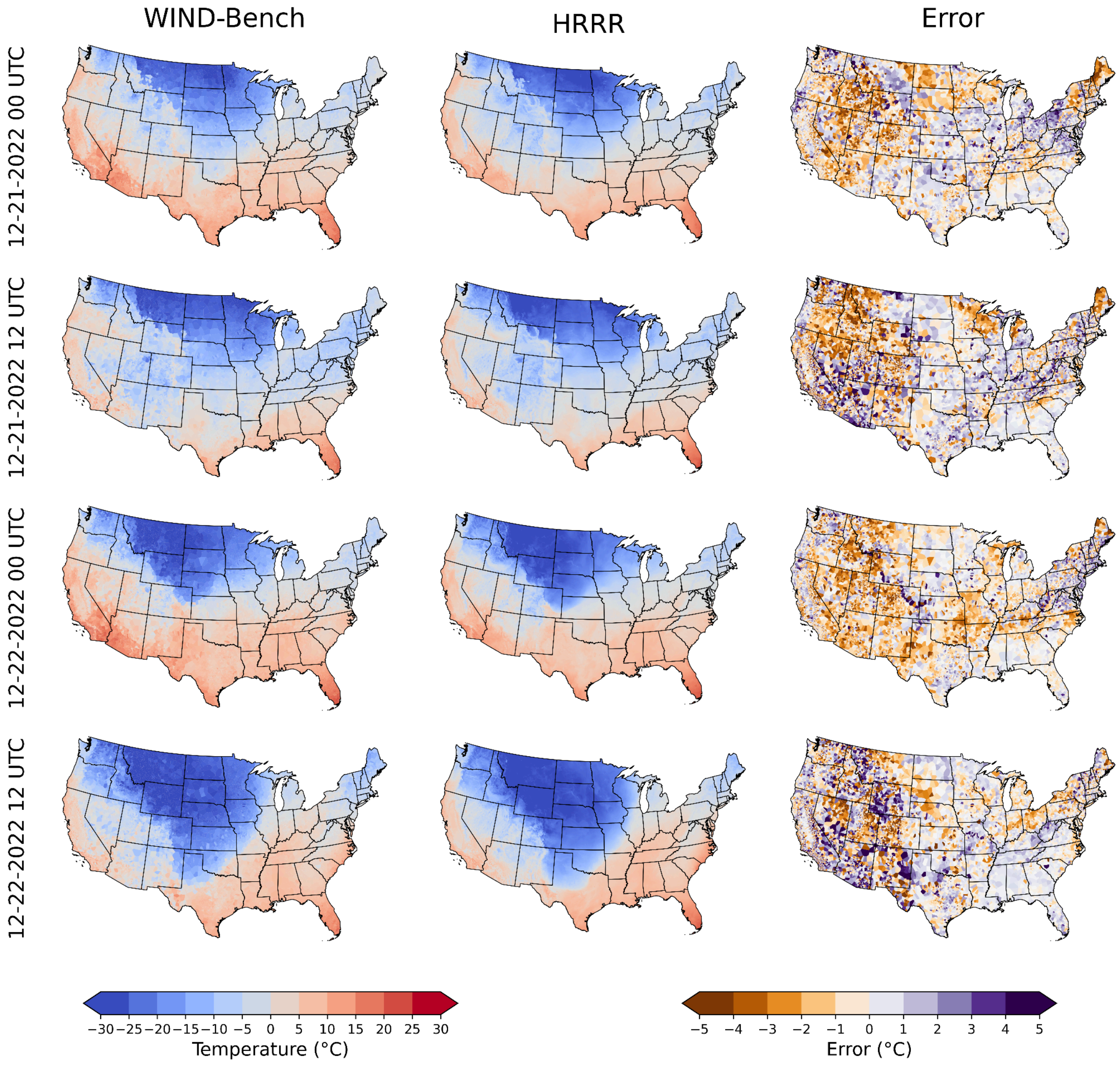

**Figure 9. Surface temperature across CONUS on December 21-22, 2022. Left column: WIND-Bench observed temperature, shown as a Voronoi diagram. Middle column: forecasted 2 m temperature from the HRRR model. Right column: temperature error, calculated as HRRR-forecasted minus MADIS-observed temperature, shown as a Voronoi diagram. Rows are time steps 12 hours apart starting at 12-21-2022 00 UTC.**

# Usage Notes

WIND-Bench is intended to support a range of scientific applications including evaluating weather models, training machine learning models, and analyzing past weather events. WIND-Bench is likely not appropriate for studying hurricanes and tornados, as it utilizes an upper wind speed threshold of 50 m $s^{-1}$ based on the apparent lack of realistic wind observations exceeding this threshold (for example, see the Section on Technical Validation). To the knowledge of the authors, the MADIS wind observations have not been validated for these types of events and it is possible that sensor hardware failures prevent high quality observations during these events. Users should be aware that although the dataset is composed of observations, erroneous data is still possible with both false positive and false negative values and further post-processing may be necessary in some cases. Station placement, sensor calibration, and other hardware failures can still affect data quality after the quality control process. Inconsistent sensor protocols and a lack of station documentation for all networks introduce uncertainty into the dataset (see Supplemental Table S1). Users should also be aware of possible selection bias introduced during the quality control process since some sensors are less reliable during extreme storms.

# Data Availability

The WIND-Bench dataset is available for public download on the Open Energy Data Initiative (OEDI) at the following URL: https://data.openei.org/submissions/8729

# Code Availability

The Python code used to download, process, and quality control the MADIS observational data is available in an open-source GitHub repository (https://github.com/NatLabRockies/madis).

# Acknowledgments and Funding

This work was authored in part by the National Laboratory of the Rockies for the U.S. Department of Energy (DOE) operated under Contract No. DE-AC36-08GO28308. Funding for K.B., G.B., B.B., L.N., and L.V. provided by the NSF CO-WY ASCEND Engine, grant NSF-2315760. E.W. was supported by a NOAA cooperative agreement for the Cooperative Institute for Research in the Atmosphere (NA24OARX432C0007), the Bipartisan Infrastructure Law (BIL) Provision 5, Fire 1 (NA23OAR40504181). This research was performed using computational resources sponsored by the U.S. Department of Energy's Office of Critical Minerals and Energy Innovation and located at the National Laboratory of the Rockies. The views expressed in the article do not necessarily represent the views of the DOE, NOAA, the Department of Commerce, or the U.S. Government. The U.S. Government retains, and the publisher, by accepting the article for publication, acknowledges that the U.S. Government retains a nonexclusive, paid-up, irrevocable, worldwide license to publish or reproduce the published form of this work, or allow others to do so, for U.S. Government purposes.

# Author Contributions

K.B. was responsible for developing quality-control algorithms, analyzing data, data visualization, and drafting the manuscript. G.B. contributed to the development of the quality-control algorithm, project conceptualization, manuscript review, and supervision of this project. B.B produced the data download and processing pipeline used to obtain observational data. L.N., D.T., A.B., and E.W. contributed to research development and manuscript review. L.V. contributed to the review of the manuscript.

# Supplemental

**Table S1. Data providers, station counts, and sensor height and data averaging intervals when available. NA signifies that documentation could not be found or there is no standard (e.g., as in the case of network aggregators).**

| Data Provider | Number of Stations | Sensor Height (above ground level) | Data Averaging | Notes | Reference |
|---|---|---|---|---|---|
| MesoWest | 14970 | Varies | Varies | Data aggregator hosted by the University of Utah | https://mesowest.utah.edu/ |
| APRSWXNET | 14922 | Varies<br>**wind:** 10m (recommended)<br>**temp:** 1.5m (recommended) | **temp:** 5-minute average<br>**ws:** 2-minute average of 5 second data<br>**wg:** maximum instantaneous value in 10-minutes before valid time | The Citizen Weather Observer Program is operated by volunteers, and sensor set ups vary, with most sensors located in urban areas. | https://www.weather.gov/media/epz/mesonet/CWOP-OfficialGuide.pdf |
| RAWS | 5114 | **temp:** 1.2-2.4m<br>**wind:** 6.1m | NA | Remote Automated Weather Stations for fire weather. | https://raws.nifc.gov/sites/default/files/inline-files/NWCG%20Standards%20for%20Fire%20Weather%20Stations_2019_1.pdf |
| METAR | 2140 | **temp:** 1.25-2m<br>**wind:** 8-10m | **temp:** 1-minute average from 6 samples per minute.<br>**ws:** 2-minute average of 5-second average<br>**wg:** Reported when 5-second average exceeds 2-minute average by 5 knots and peak gust exceeds 2 minute average by 10 knots. | Meteorological Aerodrome Report, weather stations primarily located at airports and used for pilots. This includes data from the widely used Automated Surface Observing System (ASOS) and Automated Weather Observing System (AWOS) networks. | https://www.weather.gov/media/asos/aum-toc.pdf |
| HADS | 2127 | Varies | Varies | The Hydrometeorological Automated Data System is a data aggregator, data acquisition, and distribution system. | https://hads.ncep.noaa.gov/ |
| NonFedAWOS | 584 | **temp:** 1.5-2m<br>**wind:** 10m | NA | Non-federal ASOS stations, sensor guidance provided by the | https://www.faa.gov/documentLibrary/media/ |

| Data Provider | Number of Stations | Sensor Height (above ground level) | Data Averaging | Notes | Reference |
|---|---|---|---|---|---|
| | | | | Federal Aviation Administration. | Order/JO_6560_20C.pdf |
| LCRA | 233 | NA | NA | Lower Colorado River Authority Hydromet. Primarily streamflow network with temperature. No wind. | https://hydromet.lcra.org/ |
| | | | | | |
| OHDOT | 167 | **temp:** 1.5-2m (recommended)<br>**wind:** 10m (recommended) | NA | Ohio Department of Transportation (OHDOT) Road Weather Information System (RWIS). Collection protocol is not publicly documented but may adhere to RWIS guidance. | https://www.transportation.ohio.gov/travel/driving/its/road-weather-information-system-rwis-and-weather-management<br>RWIS guidance: https://ops.fhwa.dot.gov/publications/ess05/ess0504.htm |
| ITD | 130 | **temp:** 1.5-2m (recommended)<br>**wind:** 10m (recommended) | **ws:** yes | Idaho Transportation Department (ITD) Road Weather Information System. | https://511.idaho.gov/#:Alerts<br>RWIS guidance: https://ops.fhwa.dot.gov/publications/ess05/ess0504.htm |
| WT-Meso | 122 | **temp:** 1.5m & 2m<br>**wind:** 2m & 10m | **ws:** at 10-m sensor reported as 1-min average<br>**wg:** peak 3-second gust | West Texas Mesonet records observations every minute for wind speed, direction, temperature, and relative humidity. | https://www.depts.ttu.edu/nwi/research/facilities/wtm/index.php |
| MAP | 110 | NA | NA | Multi-Agency Profiler, aggregator of sub-providers. | https://madis.ncep.noaa.gov/map_providers.shtml |
| MNDOT | 97 | **temp:** 1.5-2m (recommended)<br>**wind:** 10m (recommended) | NA | Minnesota Department of Transportation (MnDOT) Road Weather Information System. | https://www.dot.state.mn.us/maintenance/faq.html<br>RWIS guidance: https://ops.fhwa.dot.gov/publications/ess05/ess0504.htm |
| NOS-NWLON | 91 | **temp:** NA<br>**wind:** 8 - 10 m | NA | NOAA National Water Level Observation Network. Primarily a water level network. Measurements reported every 6 minutes. | https://tidesandcurrents.noaa.gov/publications/NOAA_Technical_Report_NOS_COOPS_026.pdf |

| Data Provider | Number of Stations | Sensor Height (above ground level) | Data Averaging | Notes | Reference |
|---|---|---|---|---|---|
| VADOT | 90 | **temp:** 1.5-2m (recommended)<br>**wind:** 10m (recommended) | NA | Virginia Department of Transportation (VDOT) Road Weather Information System. | https://vtrc.virginia.gov/media/vtrc/vtrc-pdf/vtrc-pdf/98-r21.pdf<br>RWIS guidance: https://ops.fhwa.dot.gov/publications/ess05/ess0504.htm |
| CoAgMet | 83 | **temp:** 1.5 m<br>**wind:** 2 m or 3; 10 m at select stations | NA | Colorado Agricultural Meteorological Network (Colorado Climate Center). | https://coagmet.colostate.edu/station_description.php |
| CC-ECONet | 71 | **temp:** 2 m<br>**wind:** 2 m, 6 m, and 10 m | **ws:** 1-minute average<br>**wg:** max 5-second wind speed over 1-minute period | Coastal Carolina Environment and Climate Observing Network. All variables except wind speed, direction, and gust are measured at 1-minute intervals. | https://econet.climate.ncsu.edu/about/<br>https://journals.ametsoc.org/view/journals/atot/40/6/JTECH-D-22-0079.1.xml |
| CAIC | 71 | Varies | Varies | Colorado Avalanche Information Center. Data provider and aggregator. | https://avalanche.state.co.us/weather/weather-stations |
| NJWxNet | 56 | Varies | Varies | The Rutgers New Jersey Weather Network. Data aggregator; NJSCO is the primary station operator. | https://www.njweather.org/njwxnet |
| DEOS | 55 | NA | NA | Delaware Environmental Observing System. | https://www.deos.udel.edu/about/ |
| KSDOT | 54 | **temp:** 1.5-2 m (recommended)<br>**wind:** 10 m (recommended) | NA | Kansas Department of Transportation Road Weather Information System. | https://rwis.ksdot.gov/<br>RWIS guidance: https://ops.fhwa.dot.gov/publications/ess05/ess0504.htm |
| IADOT | 50 | **temp:** 1.5-2 m (recommended)<br>**wind:** 10 m (recommended) | NA | Iowa Department of Transportation Road Weather Information System. | https://data.iowadot.gov/datasets/road-weather-information-system-rwis-surface-data/about<br>RWIS guidance: https://ops.fhwa.dot.gov/publications/ess05/ess0504.htm |
| NDDOT | 47 | **temp:** 1.5-2 m (recommended)<br>**wind:** 10 m (recommended) | NA | North Dakota Department of Transportation Road | https://www.dot.nd.gov/dot/rwis/<br>RWIS guidance: https://ops.fhwa.dot.go |

| Data Provider | Number of Stations | Sensor Height (above ground level) | Data Averaging | Notes | Reference |
|---|---|---|---|---|---|
| | | | | Weather Information System. | v/publications/ess05/ess0504.htm |
| MADOT | 40 | NA | NA | Massachusetts Department of Transportation. | ttps://www.mass.gov/orgs/massachusetts-department-of-transportation |
| NC-ECONet | 40 | **temp:** 2 m<br>**wind:** 10 m | NA | North Carolina Environment and Climate Observing Network. Measurements taken every minute. | https://climate.ncsu.edu/blog/2013/07/update-from-the-field-whats-happening-with-the-econet/ |
| KYTC-RWIS | 39 | **temp:** 1.5-2 m (recommended)<br>**wind:** 10 m (recommended) | NA | Kentucky Transportation Cabinet Road Weather Information System. | https://datastudio.google.com/reporting/09527215-5823-49ed-9c0b-7e5a16ad3780/page/uSEMF<br>RWIS guidance: https://ops.fhwa.dot.gov/publications/ess05/ess0504.htm |
| CA-Hydro | 38 | Varies | Varies | California Hydrometeorologic Networks. Data aggregator. | https://archive.eol.ucar.edu/projects/hydrometnet/california/ |
| SFWMD | 37 | NA | NA | South Florida Water Management District. | https://www.sfwmd.gov/sites/default/files/documents/site_status_report_user_guide.pdf |
| FL-Meso | 36 | **temp:** 1.8 m<br>**wind:** 9 m | NA | Florida Mesonet/ Florida Automated Weather Network (FAWN). Measurements are taken every 15 minutes. | https://gardeningsolutions.ifas.ufl.edu/care/weather/fawn/<br>https://arxiv.org/pdf/2310.16739 |
| NOS-PORTS | 35 | NA | NA | National Ocean Service Physical Oceanographic Real-Time System. Measurements taken every 6 minutes. | ttps://tidesandcurrents.noaa.gov/ports.html<br>https://www.tandfonline.com/doi/full/10.1080/1755876X.2018.1545558#d1e122 |
| MODOT | 35 | **temp:** 1.5-2 m (recommended)<br>**wind:** 10 m (recommended) | NA | Missouri Department of Transportation. | https://www.modot.org/sites/default/files/documents/general_services/central_office_bidding/attachment1-modotrwislocationandequipmentdetails.pdf |

| Data Provider | Number of Stations | Sensor Height (above ground level) | Data Averaging | Notes | Reference |
|---|---|---|---|---|---|
| | | | | | RWIS guidance: https://ops.fhwa.dot.gov/publications/ess05/ess0504.htm |
| VTDOT | 33 | **temp:** 1.5-2 m (recommended) **wind:** 10 m (recommended) | NA | Vermont Agency of Transportation Road Weather Information System. | https://vtrans.vermont.gov/operations/rwis RWIS guidance: https://ops.fhwa.dot.gov/publications/ess05/ess0504.htm |
| ARLFRD | 33 | **temp:** 2 m **wind:** 15 m | **ws:** 5 minute average of 1 second observations. **temp:** 5 minute average of 1 second observations. | NOAA Air Resources Laboratory/Field Research Division. Provides meteorological support to the Idaho National Laboratory. | https://inl.gov/content/uploads/2023/07/Meteorological-Monitoring_2021.pdf |
| MOComAgNet | 32 | NA | NA | University of Missouri Extension Commercial Agriculture Automated Weather Network. Some stations have 3 m towers. | http://agebb.missouri.edu/weather/stations/in1s.htm |
| DEDOT | 26 | **temp:** 1.5-2 m (recommended) **wind:** 10 m (recommended) | NA | Delaware Department of Transportation Road Weather Information System. | https://deldot.gov/map/?tab=RoadwayWeather RWIS guidance: https://ops.fhwa.dot.gov/publications/ess05/ess0504.htm |
| IEM | 24 | Varies | Varies | Iowa Environmental Mesonet, data aggregator. | https://mesonet.agron.iastate.edu/info/iem.php |
| NRCS | 23 | NA | NA | Natural Resources Conservation Service. Network unclear. | NA |
| UDFCD | 22 | NA | NA | Denver Urban Drainage and Flood Control District/ The Mile High Flood District | https://alert5.udfcd.org/resources/ |
| NHDOT | 21 | **temp:** 1.5-2 m (recommended) **wind:** 10 m (recommended) | NA | New Hampshire Department of Transportation, | https://www.dot.nh.gov/ RWIS guidance: https://ops.fhwa.dot.gov/publications/ess05/ess0504.htm |

| Data Provider | Number of Stations | Sensor Height (above ground level) | Data Averaging | Notes | Reference |
|---|---|---|---|---|---|
| CRN2-al | 16 | NA | NA | University of Alabama, Huntsville, national mesonet provider. | https://madis.ncep.noaa.gov/madis_recent.shtml |
| MEDOT | 11 | **temp:** 1.5-2 m (recommended)<br>**wind:** 10 m (recommended) | NA | Main Department of Transportation Road Weather Information System. | https://www.maine.gov/dot/<br>RWIS guidance: https://ops.fhwa.dot.gov/publications/ess05/ess0504.htm |
| INTERNET | 7 | Varies | Varies | Boulder WFO Miscellaneous, data aggregator. | NA |
| SURFRAD | 7 | NA | NA | NOAA/OAR/ESRL/GDM (Surface Radiation) Network. | https://www.ncei.noaa.gov/access/metadata/landing-page/bin/iso?id=gov.noaa.ncdc:C00540;view=iso |
| CO_E-470 | 7 | NA | NA | Colorado E-470 Public Highway Authority weather stations. | https://e470.com/ |
| MQT_Meso | 3 | NA | NA | Marquette, MI WFO Miscellaneous | NA |